\documentclass[aps,prl,twocolumn,groupedaddress,superscriptaddress,nofootinbib,floatfix,colorlinks,bookmarks=false,citecolor=darkspringgreen,linkcolor=darkolivegreen,urlcolor=navyblue]{revtex4-1}

\usepackage{graphicx}
\usepackage{amsfonts}
\usepackage{amssymb}
\usepackage{amsmath}
\usepackage{dcolumn}
\usepackage[normalem]{ulem}
\usepackage{enumerate}
\usepackage{bbold}
\usepackage{cancel}
\usepackage{mathtools}
\usepackage{siunitx}
\usepackage{braket}
\usepackage{amsmath}
\usepackage{mdframed}
\usepackage{dsfont}
\usepackage{float}
\usepackage[caption=false]{subfig}
\usepackage{soul}
\usepackage{orcidlink}
\usepackage{placeins}
\usepackage{stackrel}

\definecolor{darkgreen}{RGB}{0, 159, 117}
\definecolor{navyblue}{rgb}{0.0, 0.0, 0.5}
\definecolor{darkolivegreen}{rgb}{0.33, 0.42, 0.18}
\definecolor{darktangerine}{rgb}{1.0, 0.66, 0.07}
\definecolor{darkspringgreen}{rgb}{0.09, 0.45, 0.27}

\usepackage{mathrsfs}

\newcommand{\up}{\uparrow}
\newcommand{\down}{\downarrow}

\newcommand{\papertitle}{Entanglement capacity of complex networks from quantum walks}

\begin{document}
\title{\papertitle}

\author{Pravy Prerana}
\email{preranap@uni.coventry.ac.uk}
\affiliation{
Centre for Fluid and Complex Systems, Coventry University, Coventry, CV1 2TT, United Kingdom}

\author{Sascha Wald}
\email{sascha.wald@coventry.ac.uk}
\affiliation{
Centre for Fluid and Complex Systems, Coventry University, Coventry, CV1 2TT, United Kingdom}

\begin{abstract}

Discrete-time quantum walks provide a natural framework for quantum transport on complex networks. 
On regular structures, coin-walker entanglement has been widely used to characterize quantum transport and to support quantum algorithmic protocols.
However, this notion relies on a fixed Hilbert space factorization separating coin and position and is therefore not directly applicable to more complex, irregular structures. 
Here we introduce an entanglement measure for general networks based on a bipartition that assigns each node two roles, acting as both a source and a target.
The resulting bipartition defines the {\it source-target entanglement}, a measure for general networks, motivated by coin–walker entanglement.
We show that the connectivity of the network imposes an upper bound on this entanglement and identify graph matchings as the underlying structure governing entanglement generation.
We further illustrate that in random graphs improving graph connectivity reduces the attainable entanglement, establishing a structure-dependent constraint on quantum correlations.
\end{abstract}

\maketitle

{\bf Introduction.---}Discrete-time quantum walks (DTQWs) are paradigmatic models for coherent quantum transport, in which a `quantum coin toss' dictates the dynamics of the quantum walker~\cite{Aha93, Kem03, Aha01,Venegas12,Xia20, Kadian2021}.
Not only can these walks model the spreading of quantum particles~\cite{Kar09,Yao23} but they are also universal computational primitives~\cite{Childs09,Lovett10} and provide a natural framework for describing active quantum agents~\cite{Yam24,gipou26} in the emerging field of active quantum matter~\cite{Kha24,Nado25,antonov2025,antonov2026}. 
What sets DTQWs apart from their classical counterparts is the coin–walker entanglement generated during the evolution, which can give rise to their characteristic ballistic spreading~\cite{Vieira13,Vieira14}.
This entanglement has emerged as a resource in quantum algorithms~\cite{Sanchez12,Xia20,Shenvi03,Childs04,Tulsi08,Xu25}, quantum state engineering~\cite{Inno17,Gio19}, quantum communication protocols~\cite{Sri20,Gio21,Panda23,Chawla23,Das25} and the simulation of topological phenomena~\cite{Kita12,Wang19,Panda25,Panda26,Amri19}. 

DTQWs have been realized on various platforms~\cite{Man14}, including synthetic photonic lattices~\cite{Knight03, Knight03a,Banuls06}, neutral atom traps~\cite{Eck05,Chan06} and quantum circuits~\cite{Koi05} with notable extensions to DTQWs on complex networks~\cite{Hin07,Dou09,Loke11}.
Although quantum walks on graphs have attracted significant attention~\cite{Xiao21, Boet21, Sanchez12, Wald21, Aca20,Xu21,Krovi06,Krovi07} and despite its centrality, entanglement in DTQWs has been studied primarily on regular structures where the coin and the walker space are separable~\cite{Port18}.
However, this has left quantum correlations in walks on irregular graphs largely unexplored, raising a fundamental question: to what extent does the underlying graph structure govern the entanglement a quantum walk can generate?

Here we introduce a graph-level entanglement measure for DTQWs, and demonstrate that it encodes structural information about the underlying graph. 
\begin{figure}
    \centering
    \includegraphics[width=\columnwidth]{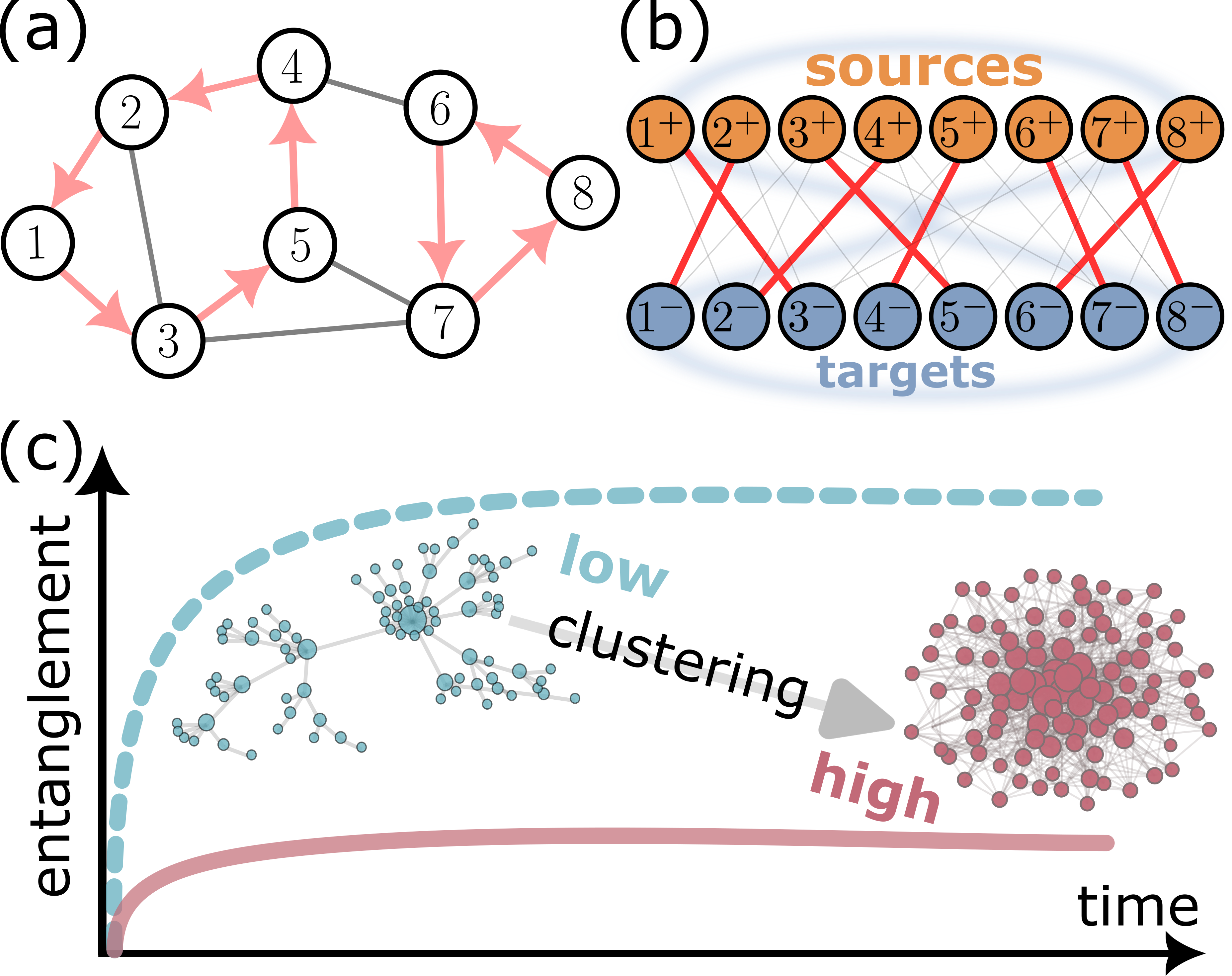}
    \caption{{\bf Illustration DTQW states and their source-target entanglement}: (a) A generic graph G, composed of eight nodes and twelve edges. 
    Arrows highlight edges that contribute to a DTQW state defined on the symmetric digraph ${\cal G}$, cf. Eq.~(\ref{eq:state}).
     (b) Bipartite double cover of the graph $G$ from panel (a).
     Each node $n$ is split into a source and target ($n^{\pm}$) and states $\ket{n\to m}$ are mapped to $\ket{n^+}\otimes \ket{m^-}$.
     The bold red edges form a maximum matching of $B(G)$ and correspond to the DTQW state illustrated in panel (a).
     (c)  Source-target entanglement generated during the time evolution of a DTQW on networks with different structural properties.
    Low clustering, hub dominated structures yield higher entanglement compared to highly clustered networks where the entanglement becomes significantly suppressed.}
    \label{fig:Fig1}
\end{figure}
To this end, we consider the bipartite double cover of the graph, which splits each node $n$ into a source and a target node $n \mapsto (n^+,n^-)$, see Fig.~\ref{fig:Fig1}.
The source-target division takes over the role of the traditional coin-walker division as it quantifies correlations between the walker's position and its orientation.
However, while coin-walker entanglement quantifies global directional alignment between different components of the quantum state,
source-target entanglement determines independent coherent transport of distinct components of the quantum state.

In this work we develop the intricate connection between source-target entanglement and the graph connectivity pattern and establish tight bounds on the entanglement capacity of complex networks.
We investigate how DTQWs can generate source-target entanglement and discuss the role of the network.
In particular we find that maximally entangled states are directly connected to graph matchings,  and general improvements in the graph connectivity yield diminished entanglement, see Fig.~\ref{fig:Fig1}.
These results directly quantify the coherence of DTQWs on generic networks, and the network's ability to support quantum correlations.\\

{\bf Quantum walks on graphs.---}We consider DTQWs on generic graphs $G$ with the set of nodes and edges $N,E$ respectively, see Fig.~\ref{fig:Fig1}(a).
DTQWs are naturally defined on the symmetric digraph $\mathcal{G}$ obtained by replacing each edge $(n,m)\in E$ with two arcs $n\to m$ and $m\to n$~\cite{Port18}.
The Hilbert space of the walk is thus spanned by the arc states and a DTQW state can be expressed as
\begin{align}
\ket{\psi}=\sum_{n\to m\in A}\psi_{n\to m}\ket{n\to m}
\label{eq:state}
\end{align}
with the set of all arcs $A$.
The time evolution is generated by the iterative application of a coin operator $\mathscr{C}$ followed by the flip-flop shift operator $\mathscr{S}$, i.e. $\ket{\psi(t+1)}=\mathscr{S} \mathscr{C}\ket{\psi(t)}$.
The coin operation generates local superpositions of arc states with the same tail while the flip-flop shift operator inverts arc states by exchanging head and tail $\mathscr{S}\ket{n\to m} = \ket{m\to n}$.
In irregular graphs, the coin space cannot be separated as a tensor product from the walker space due to non-uniform node degrees.
Therefore, coin-walker entanglement on these complex structures cannot be quantified in the same way as in the regular case.

To quantify entanglement on general graphs $G$, we introduce the corresponding bipartite double cover $B(G)$, see Fig.~\ref{fig:Fig1}(b).
Each node $n\in N$ is split into a source node $n^+$ and a target node $n^-$ while each edge $(n,m)$ induces the edges $(n^+,m^-)$ and $(m^+, n^-)$ in $B(G)$.
Since there is a canonical one-to-one correspondence between the bipartite double cover and the symmetric digraph, walker states such as~(\ref{eq:state}) can be embedded into the Hilbert space of the bipartite double cover, viz.
\begin{align}
\ket{\Psi}=\sum_{n,m\in N}\Psi_{nm}\ket{n^+}\otimes\ket{m^-}.
\label{eq:state_bdc}
\end{align}
The coefficient matrix is given by $\Psi_{nm} = \psi_{n\to m}$ whenever $n\to m \in A$ and zero otherwise.
Similar descriptions of QWs in terms of the bipartite double cover have been used in Szegedy's formulation \cite{Wong2017,Port2016,Port2017,Xu21a}.

The representation~(\ref{eq:state_bdc}) in terms of $B(G)$ naturally induces a unique source–target bipartition of the walker state.
We quantify this  {\it source-target entanglement} using the von Neumann entropy
\begin{align}
S_{\rm st}=-\sum_{j=1}^{\ell}\lambda_j^2\log\lambda_j^2,
\end{align}
where $\lambda_j$ are the Schmidt coefficients of $\ket{\Psi}$, i.e. $\ket{\Psi}=\sum_{j=1}^\ell \lambda_j \ket{\alpha_j^+}\otimes \ket{\beta_j^-}$ \cite{Nielsen10,Sper11}.
Since the graph structure restricts the support of the amplitude matrix $\Psi$, this representation naturally relates the entanglement structure of the walker state to the sparsity pattern imposed by the graph.
In particular, the Schmidt rank of a state is constrained by which pairs of nodes are connected by arcs.

\begin{figure}
    \centering
    \includegraphics[width=\columnwidth]{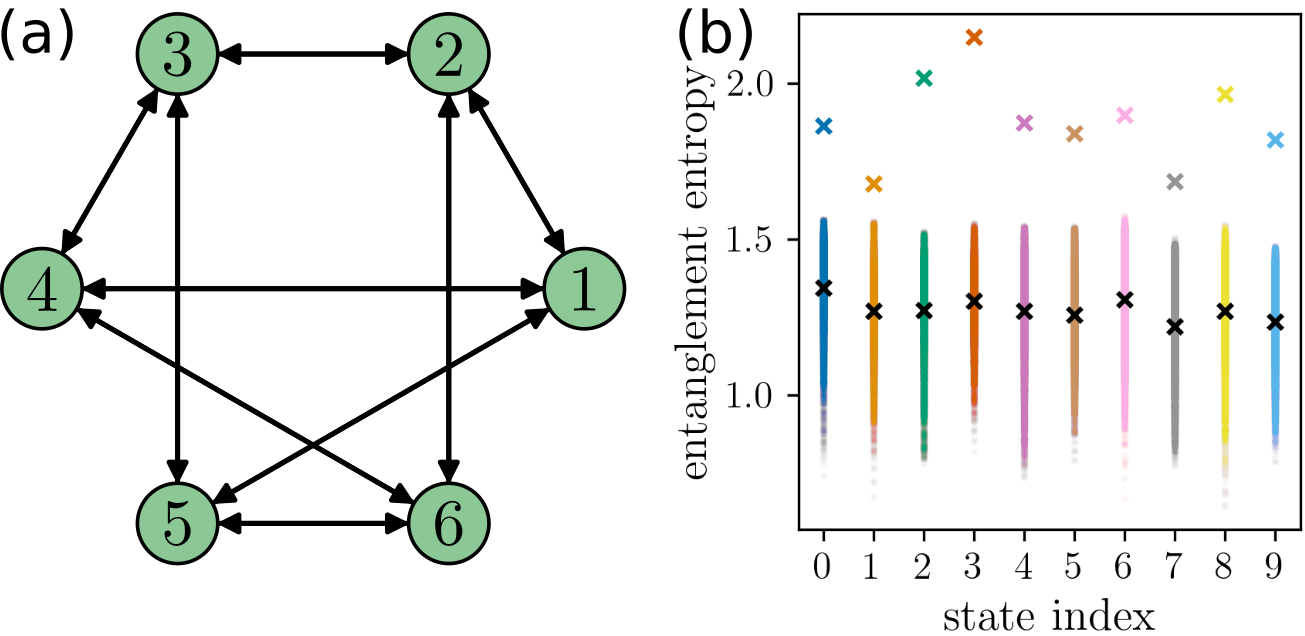}
    \caption{{\bf Entanglement of Haar random states on regular graph.}
    We consider the three-regular symmetric digraph in panel (a) and generate Haar random states in the arc basis.
    In panel (b) we evaluate the source-target entanglement of ten states (colored cross) and compare it to the coin-walker entanglement for all $(3!)^6 = 46656$ distinct coin assignments (scatter lines, black crosses indicating averages).
    }
    \label{fig:st-vs-cw}
\end{figure}

The source-target bipartition replaces the traditional coin-walker division as a unique quantifier of position-orientation correlations of the walker.
Even when a tensor-product structure between coin and position exists, as in regular graphs, the specific identification between coin states and outgoing edges is not unique.
As illustrated in Fig.~\ref{fig:st-vs-cw}, different coin assignments can lead to substantially different values of the coin–walker entanglement.
While one may impose a globally consistent labeling via an edge coloring, this is not required for defining a DTQW, and standard constructions such as the Hadamard walk~\cite{Nayak00,Kem03} rely only on local assignments, see Supplemental Material for details~\cite{SM}.
Consequently, the three-regular graph in Fig.~\ref{fig:st-vs-cw}(a) admits $(3!)^6$ distinct local coin assignments with a wide spread of coin-walker entanglement, see Fig.~\ref{fig:st-vs-cw}(b).
In~\cite{SM} we also compare the source-target entanglement with the coin-walker entanglement in the case of the Hadamard walk, establishing that their behavior is distinct.\\

{\bf Entanglement capacity of graphs.---}It is the central objective of this work to quantify the source-target entanglement of walker states on complex networks.
We note that superposition of states that share the same source or target are not entangled.
The source-target entanglement thus quantifies the extent to which the walker state is spread over distant parts of the graph.
This spread can be quantified with {\it graph matchings}, which are collections of distinct graph edges with no shared nodes.
Here we consider matchings of the bipartite double cover $B(G)$ that can be formulated in terms of arcs of the symmetric digraph ${\cal G}$.
A matching  $ M \subset A$ is a set of arcs on the digraph such that no two arcs share the same head or the same tail.
The cardinality $|M|$ is called the size of the matching.
Matchings of largest possible size are called {\it maximum matchings} and matchings that match all nodes are called {\it perfect matchings}.
The matching structure of the underlying graph allows us to introduce the graph's entanglement capacity.
This capacity quantifies the maximal amount of source-target entanglement a walker state can attain on a graph according to the following theorem.

\vspace{.25cm}

{\it
\textit{Theorem.---}For a graph $G(N,E)$, we consider a quantum walker state $\ket{\Psi}$ on the corresponding symmetric digraph ${\cal G}(N,A)$, i.e. $\ket{\psi} = \sum_{n\to m \in A} \psi_{n\to m} \ket{n\to m}$.
This state induces a set of support edges  $E_{\ket{\psi}} =\{ (n^+,m^-) |\psi_{n\to m} \neq 0  \}$ on the bipartite double cover $B(G)$.
Let $s$ be the size of the largest matching of $B(G)$ that is fully contained in $E_{\ket{\psi}}$.
The source-target entanglement is then bounded by $S_{\rm st} [\ket{\psi}] \leq \log s$.}

\vspace{.1cm}

\textit{Proof.---}
Embedding the walker state into the bipartite double cover maps the amplitudes onto the matrix $\Psi$, defined in Eq.~(\ref{eq:state_bdc}).
We perform a singular value decomposition on $\Psi = U \Lambda V^\dagger$ which reveals that the matrix rank of $\Psi$ is the Schmidt rank of $\ket{\Psi} = \sum_i \Lambda_i \sum_n U_{ni}\ket{n^+} \otimes \sum_m V_{im}^\dagger\ket{m^-}$.
Since arcs sharing the same source or target occupy the same row or column, only entries corresponding to a matching can contribute independently to the matrix rank.
Hence, the rank of $\Psi$ is bounded by the size $s$ of the largest matching contained in $E_{\ket{\Psi}}$, which implies $S_{\rm st}[\ket{\psi}]\le \log s$.
\hfill $\square$

\vspace{.25cm}

The bound given in the theorem is also tight:
For any matching $M$ we can construct a corresponding matching state
\begin{align}
\ket{M(\phi)}=\frac{1}{\sqrt{|M|}}\sum_{\nu\in M}e^{i\phi_\nu}\ket{\nu},
\label{eq:mstate}
\end{align}
whose amplitude matrix $\Psi$ in the bipartite double cover representation is a (complex) permutation submatrix of size $|M|$.
Consequently, $\text{rank}(\Psi)=|M|$ and $S_{\rm st}=\log |M|$.
This implies that the entanglement capacity of a graph is given solely by the size of the maximum matching of its bipartite double cover.
The physical interpretation of this finding is that matchings identify sets of arcs along which quantum amplitudes can be supported independently, providing the basic resource for generating source–target entanglement.

As an example, we consider the cyclic graph with $N$ nodes.
The entanglement capacity of these graphs is always $\log N$, e.g. attained by the state $\ket{\circlearrowright}=\sum_{n=0}^{N-1} \ket{n\to (n+1)\operatorname{mod}N}/\sqrt{N}$.
For $N$ odd, apart from global change of direction and phase modulation similar to Eq.~(\ref{eq:mstate}), this is the only maximally entangled state.
However, for $N$ even, the cyclic graph itself is bipartite and each maximum matching of the graph induces a maximum matching of the bipartite double cover which is not true for $N$ odd.
Hence, for $N$ even, a second class of maximally entangled states exists that is a superposition of local cycles: $\ket{\leftrightarrow} = 1/\sqrt{N}\sum_{n=0}^{N/2-1} \ket{2n\to 2n+1} + \ket{2n+1\to 2n}$.

\begin{figure*}[t]
    \centering
    \includegraphics[width=\textwidth]{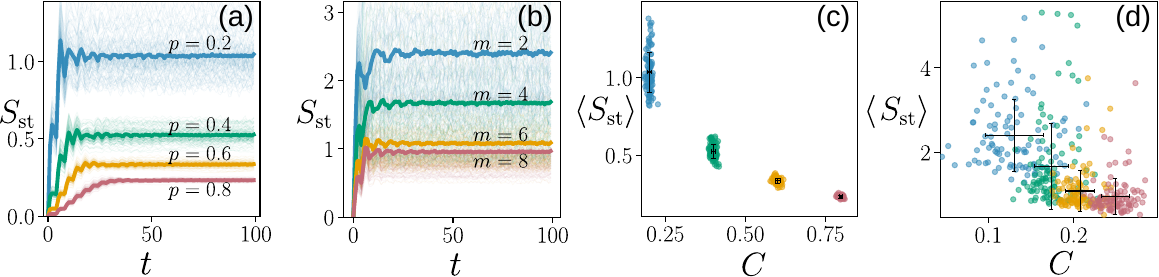}
    \caption{{\bf Source--target entanglement in random graphs.}
     Panels (a) and (b) show the time evolution of the source-target entanglement entropy $S{\rm_{st}}$ for ER and BA graphs respectively.
    Faint lines depict the entanglement entropy for 100 different graph realizations and the solid line is the average of these realizations.
    We observe that DTQWs on both network models exhibit higher entanglement at low connectivity, whereas increasing connectivity reduces the entanglement.
    Panels (c) and (d) present scatter plots of the time-averaged plateau values from panels (a) and (b) against the average clustering coefficient $C$ of each realization, for the ER and BA models, respectively.
    We again observe that increased connectivity as evidenced by higher clustering diminishes the generated entanglement.
    }
    
    \label{fig:dynamics}
\end{figure*}

The theorem allows us to directly relate the entanglement capacity of a complex network to its structural properties.
For example, it enables quantitative predictions for random network models such as Erd\H{o}s--R\'enyi (ER) graphs~\cite{Albert02}.
ER graphs are stochastic one-parameter networks with $p\in (0,1)$ indicating the probability that any two nodes in the graph are connected.
The corresponding degree distribution is binomial and ER graphs are single-scaled.
Evidently, the larger $p$ the more high degree nodes are present in the network, facilitating high connectivity and small average distance.
For ER graphs with $N$ nodes and mean degree $\bar{k}$, the expected size of the maximum matching $M^*$ is given by~\cite{Karp81,Zde06}
\begin{align}
\langle |M^*| \rangle =
\frac{N}{2}\left(
2-y-(1+\bar{k}y)e^{-\bar{k} y}
\right),
\end{align}
where $y$ is the smallest solution of the transcendental equation $y = e^{-\bar{k} e^{-\bar{k} y}}$.
Hence, we can expect that the size of the maximum matching of the bipartite double cover is larger than or equal to $2\langle |M^*| \rangle$ which is thus a lower bound for the expected entanglement capacity of an ER graph, see Supplemental Material for details~\cite{SM}. \\

{\bf Entanglement generation.---}Having quantified the entanglement capacity of generic graphs, we now consider quantum walk dynamics on random graphs.
In particular, we consider the previously introduced ER model, together with the Barab\'asi–Albert (BA) model~\cite{Albert02}, which is characterized by a scale-free degree distribution.
In the ER ensemble, edges are placed independently with probability $p$, leading to a homogeneous network structure in which the average degree acts as the primary control parameter governing connectivity~\cite{Albert02}.
By contrast, BA networks are generated through stochastic growth: each newly added node attaches to $m$ existing nodes via preferential attachment to highly connected nodes. 
This mechanism produces hubs and a heavy-tailed degree distribution.
As $m$ increases, each new node connects to more existing nodes, increasing the average degree and reducing the average shortest-path length. 
However, the network retains its scale-free, hub-dominated structure rather than becoming truly decentralized.

For each model we generate $100$ connected network-realizations with $N=100$ nodes for a variety of parameters (ER: $p = 0.2,0.4,0.6,0.8 $; BA: $m=2,4,6,8$).
We initialize the walker in a single arc state and let it evolve for $100$ time steps using the Grover coin~\cite{Port18}.
The results are depicted in Fig.~\ref{fig:dynamics}.
We see that in all cases the DTQW rapidly generates source-target entanglement before settling into a fluctuating plateau value.
For ER graphs, increasing the attachment probability $p$ leads to a systematic reduction of the generated source–target entanglement as shown in Fig.~\ref{fig:dynamics}(a).
This behavior can be understood in terms of the changes to the graph structure.
Upon increasing $p$,  the mean degree and the edge density both increase yielding improved connectivity of the network realizations.
Source-target entanglement is readily generated by components of the quantum states residing on distant parts of the graph.
Such structural separation induces correlations between source and target degrees of freedom that cannot be factorized.
However, increasing the attachment probability in ER networks hinders this mechanism and homogenizes the network resulting in diminished source-target entanglement.
We observe a similar behavior in BA networks upon increasing $m$ (see Fig.~\ref{fig:dynamics}(b)) although BA networks will remain structurally heterogeneous.

To relate the dynamical results to structural properties of the underlying random graphs, we time-average the plateau regime for each realization and compute the average clustering coefficient of the corresponding network.
The clustering coefficient quantifies the density of triangles, i.e. the fraction of neighbor pairs of a node that are themselves connected. 
High clustering therefore indicates the presence of locally closed loops and redundant short-range connections. 
Fig.~\ref{fig:dynamics} shows how the generated entanglement relates to the average clustering of ER (panel (c)) and BA graphs (panel (d)).
This reveals a clear anticorrelation: realizations with larger clustering coefficients systematically generate less source–target entanglement.
This behavior admits a natural structural interpretation:
High clustering implies that the underlying network is strongly interconnected.
As a result, amplitudes that spread from a given node can be quickly reinjected into nearby regions via short loops, promoting local interference within tightly knit neighborhoods.
This suppresses the formation of spatially separated amplitude components that would otherwise support strong source–target correlations.

In contrast, graphs with low clustering are more locally tree-like. Such structures favor branching transport into weakly connected regions, enabling partially independent amplitude components to persist. 
This structural separation enhances the generation of source–target entanglement.\\

{\bf Conclusion.---}DTQWs are paradigmatic models for coherent quantum transport and fundamental primitives for quantum computation, which has driven much of the interest they have attracted.
On regular structures, distinct properties from their classical counterparts stem largely from coin-walker entanglement.
However, on more complex, irregular structures, the coin and position Hilbert spaces do not simply factorize into a tensor product anymore such that a new quantity is required to capture correlations generated by the geometry of the underlying graph itself.

Here, we considered DTQWs on generic graphs and used a source-target description in terms of the bipartite double cover of the graph that allows to introduce an alternate entanglement measure.
We demonstrated that the source–target entanglement of a quantum walk state is tightly constrained by graph-theoretic properties: the maximal matching of the underlying graph determines its entanglement capacity and thus the largest active matching in a quantum state bounds the state's source-target entanglement.

In the case of ER graphs, the expected size of the maximum matching is established which allows to explicitly evaluate a lower bound for the expected entanglement capacity of ER models.
Furthermore, we have shown that equal weight superpositions from maximum matchings are maximally entangled states. 
These results illustrate that the source-target entanglement is intimately connected to the connectivity properties of the digraph and we further explored this connection through numerical simulations of quantum walker dynamics on random graphs:
we considered DTQWs on randomly generated realizations of ER and BA networks as paradigmatic models of scaled and scale-free graphs.
The walker dynamics generates source-target entanglement, rapidly reaching a plateau window in which the entanglement entropy oscillates.
By increasing the connectivity in both network models, we observe that the  plateau decreases monotonically, indicating that less source-target entanglement is generated.
The connectivity can also be quantified by the average clustering coefficient and both network models indeed display a decrease in the source-target entanglement upon increasing the average clustering coefficient.

The interplay between graph connectivity and entanglement generation remains largely unexplored.
In this work, we establish a framework to quantify source–target entanglement on graphs and demonstrate how it emerges from the combined effects of graph structure and quantum dynamics.
From this perspective, source–target entanglement serves as a natural diagnostic of quantum transport, capturing the build-up of nonlocal correlations between distinct nodes of the network.

Our results establish source-target entanglement generation as an intrinsic feature of quantum walks on graphs and provide a route to characterizing correlation spreading in complex networks.
This has direct implications for quantum communication protocols and networked quantum information processing, where control over correlations is essential. 
Moreover, our framework opens the door to studying how modifications of the dynamics influence entanglement generation and may enable control over localization~\cite{Jahnke08} and search processes on graphs~\cite{Childs04,Chak16,Yin23}.\\

{\bf Data availability.---}The data displayed in the
figures is available on Zenodo~\cite{zenodo}\\

{\bf Acknowledgment.---}We would like to thank Federico Carollo, Chris Hooley and Rudolf R\"omer for fruitful discussions that helped advance this work.

\bibliographystyle{apsrev4-2}
\bibliography{refs}

\clearpage

\setcounter{equation}{0}
\setcounter{figure}{0}
\setcounter{table}{0}
\makeatletter
\renewcommand{\theequation}{S\arabic{equation}}
\renewcommand{\thefigure}{S\arabic{figure}}
\FloatBarrier
\makeatletter
\onecolumngrid
\newpage
\setcounter{page}{1}
\begin{center}
{\Large SUPPLEMENTAL MATERIAL}
\end{center}
\begin{center}
\vspace{0.8cm}
{\Large \papertitle }
\end{center}
 \begin{center}
 Pravy Prerana$^{1}$ and
 Sascha Wald$^{1}$
 \end{center}
 \begin{center}
 $^1${\em Centre for Fluid and Complex Systems, Coventry University, Coventry, CV1 2TT, United Kingdom}\\
 \end{center}
\section{Coin assignment in the Hadamard walk}
\label{sec:section 1}
\begin{figure}[H]
    \centering
    \includegraphics[width=0.7\columnwidth]{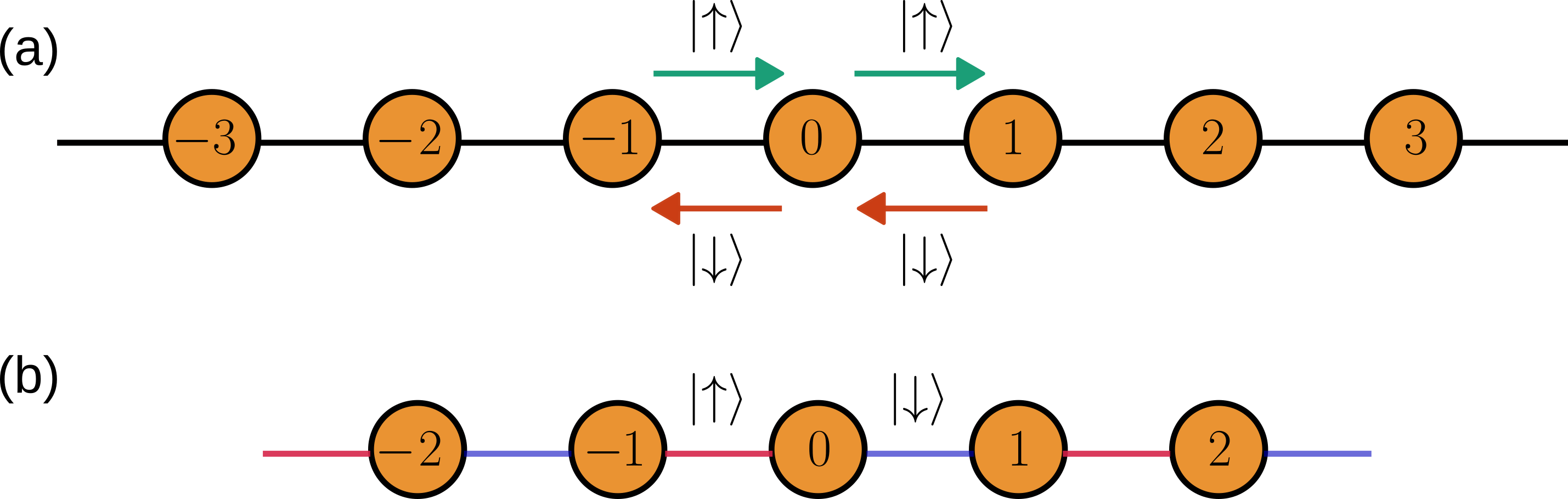}
    \caption{{\bf{Coin-walker representation for the Hadamard walk.}}
    (a) Schematic illustration of the routine coin assignment for the Hadamard walk on the infinite line.
    Here, the coin states $\ket{\up}$ and $\ket{\down}$ determine the global direction of the components, i.e. right and left.
    (b) Alternatively, the coin assignment for the Hadamard walk can also be generated from a graph coloring when the line is viewed as class 1 graph with edge chromatic number 2. 
    However, to generate the standard dynamics of the Hadamard walk, either the shift or the coin operator need to be adjusted.
    }
    \label{fig:line}
\end{figure}
In this section, we demonstrate an intrinsic ambiguity in defining a tensor product structure for the coin-walker bipartition.
In order to define a coin Hilbert space that is separable as a tensor product from the position degree of freedom, the graph must be of class 1~\cite{Port18}, i.e. all nodes must have the same degree and this degree must coincide with the graph's edge chromatic number.
In this case, each edge coloring provides a natural and globally consistent definition of the coin space.
In general, DTQWs are predominately studied on regular, periodic structures that classify as class 1 graphs.
Under such colorings the coin operators assumes a block diagonal structure, with each block being identical. 
This allows to define the bipartite state into components residing on the node space and components residing on the coin space.
Consequently, coin-walker states derived from such graph colorings have unique coin-walker entanglement but often, the coin is {\it not} explicitly derived from an edge coloring but from local coin assignments that stem from global property of the regular structure.
A typical example is the Hadamard walk on the line.
Since the line possesses a global sense of direction, i.e. left and right, the Hadamard walk is routinely described by a spin $1/2$ coin, i.e.,

\begin{align}
    \ket{\phi} = \ket{\phi_\up}\otimes \ket{\uparrow} + \ket{\phi_\down} \otimes \ket{\downarrow}\ .
\end{align}
Here, the $\ket{\up}$ component induces motion to the right, and the $\ket{\down}$ component motion to the left with the shift operator
\begin{align}
    \mathscr{S} &= \sum_x \ket{x+1,\uparrow}\bra{x,\uparrow} + \ket{x-1,\downarrow}\bra{x,\downarrow} \ .
\end{align}
The coin operator is given by the Hadamard gate $\mathscr{C} = ((\ket{\up} + \ket{\down})\bra{\up} + (\ket{\up} - \ket{\down})\bra{\down})/\sqrt{2}$.
This setup is illustrated in Fig.~\ref{fig:line}(a).
Clearly the line is two-regular and has edge chromatic number $2$. 
However, the above prescription for the Hadamard walk assigns a particular edge two distinct coin states depending on whether it is traversed leftward or rightward.
In this representation, the coin space is introduced independently of the coloring of the underlying graph.
This corresponds to a local assignment of coin states on the digraph and thus allows, in principal, to arbitrarily color the arcs.
However, when the line is viewed in terms of graph theory, the coin states can be associated with edge coloring, as shown in Fig.~\ref{fig:line}(b).
This leads us to the following unsatisfactory observation.
We consider a particular state of the Hadamard walk that can be written as  $\ket{\psi} = (\ket{0\to 1} + \ket{2\to1})/\sqrt{2}$.
In this manner, the DTQW state is well-defined with components that reside on sites $0$ and $2$ and intend to move to site $1$.
In the standard coin-walker notation this state can be written as  $\ket{\phi} = (\ket{0\up} + \ket{2\down})/\sqrt{2}$, admitting maximum coin-walker entanglement.
If we however consider the coin description given in Fig.~\ref{fig:line}(b), i.e. $\ket{\phi_{\rm CW}} = (\ket{0\up} + \ket{2\up})/\sqrt{2}$ we see that the state is fully separable.
This dependence on the choice of coin assignment highlights that coin–walker entanglement is not an intrinsic property of the underlying graph, leaving room for ambiguity in its interpretation.

\section{Comparison of coin-walker and source-target entanglement in the Hadamard Walk}

\begin{figure}[h]
    \centering
    \includegraphics[width=0.3\columnwidth]{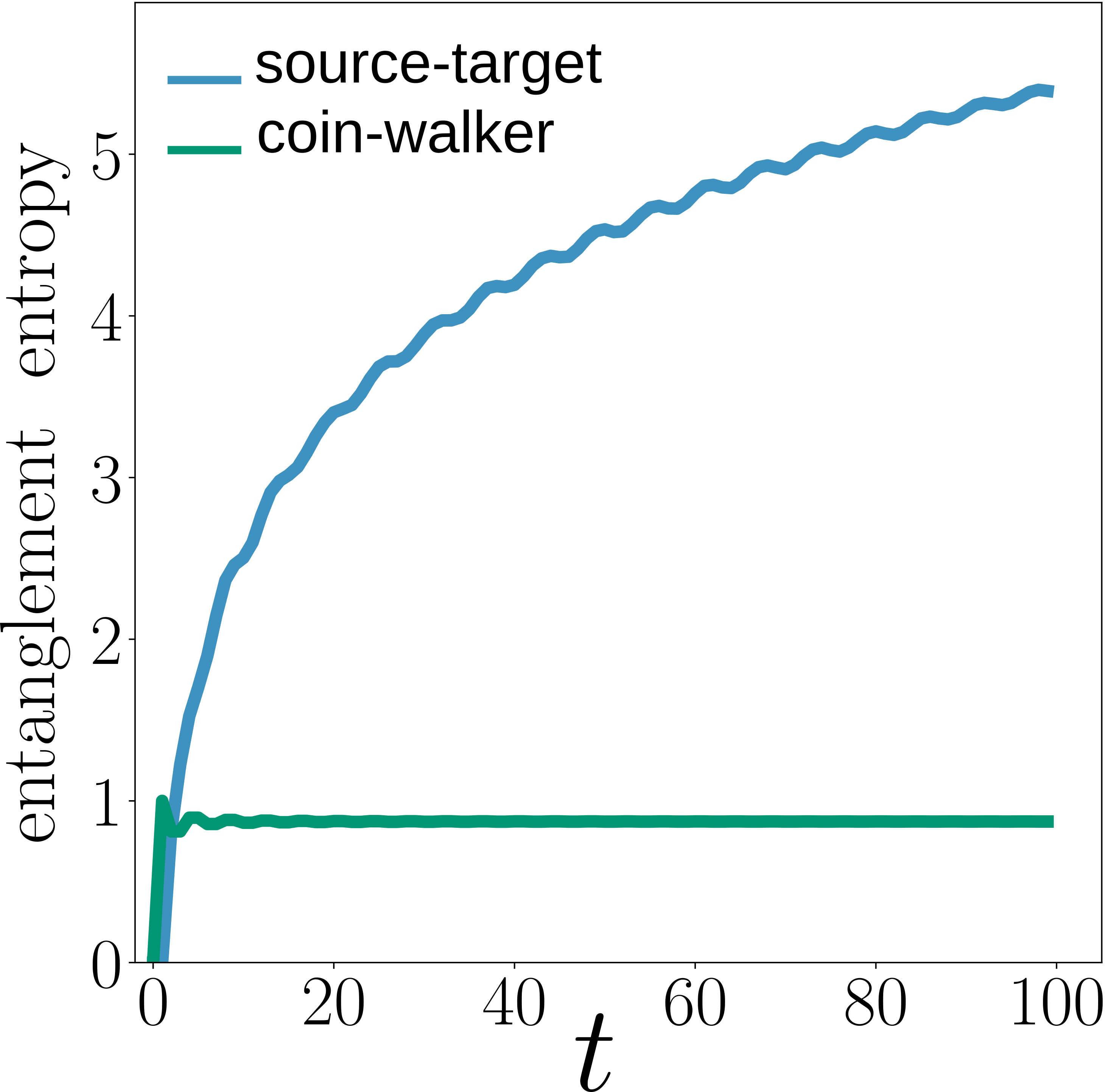}
    \caption{{\bf Coin-walker vs source target entanglement for the Hadamard walk on the infinite line.} We show a comparison between the coin-walker entanglement entropy for the Hadamard walk with the coin assignment shown in Fig.~\ref{fig:line}(a) and the source-target entanglement for the same dynamics. 
    The initial walker state considered here is $\ket{0} \otimes \frac{1}{\sqrt{2}}(\ket{\uparrow}+ i \ket{\downarrow})$ and the quantum state has been evolved for $t=100$ time steps using the usual coin and shift operator of the Hadamard walk~\cite{}.
    }
    \label{fig:cycle_EE}
\end{figure}

In this section we compare the coin-walker entanglement in the standard formulation of the Hadamard walk on the infinite line with the source-target entanglement introduced in the main text.
To evaluate the source-target entanglement, we use the identification $\ket{n \up/\down} = \ket{n\to n\pm1}$ and follow the prescription from the main text to lift the state to the bipartite double cover of the infinite line.
These two representations correspond to two different tensor-product structures and therefore, entanglement obtained from the coin–position bipartition does not generally coincide with that obtained from the source–target bipartition.
In Fig.~\ref{fig:cycle_EE} we show the time evolution of both entanglement measures for a Hadamard walk initialized at the origin of the infinite line with the initial coin state $(\ket{\up} + i \ket{\down})/\sqrt{2}$. 
We see that both quantities behave quite differently: 
Although both measures show a sharp initial rise with subsequent leveling off, not only are the scales different, also the coin-walker entanglement quickly reaches its stationary value and does not indicate any walker dynamics any more, whereas the source-target entanglement still clearly evidences the persistent spreading.

\section{Graph matching in generic graphs $G$ and their BDC $B(G)$}
\begin{figure}[t]
    \centering
    \includegraphics[width=0.5\columnwidth]{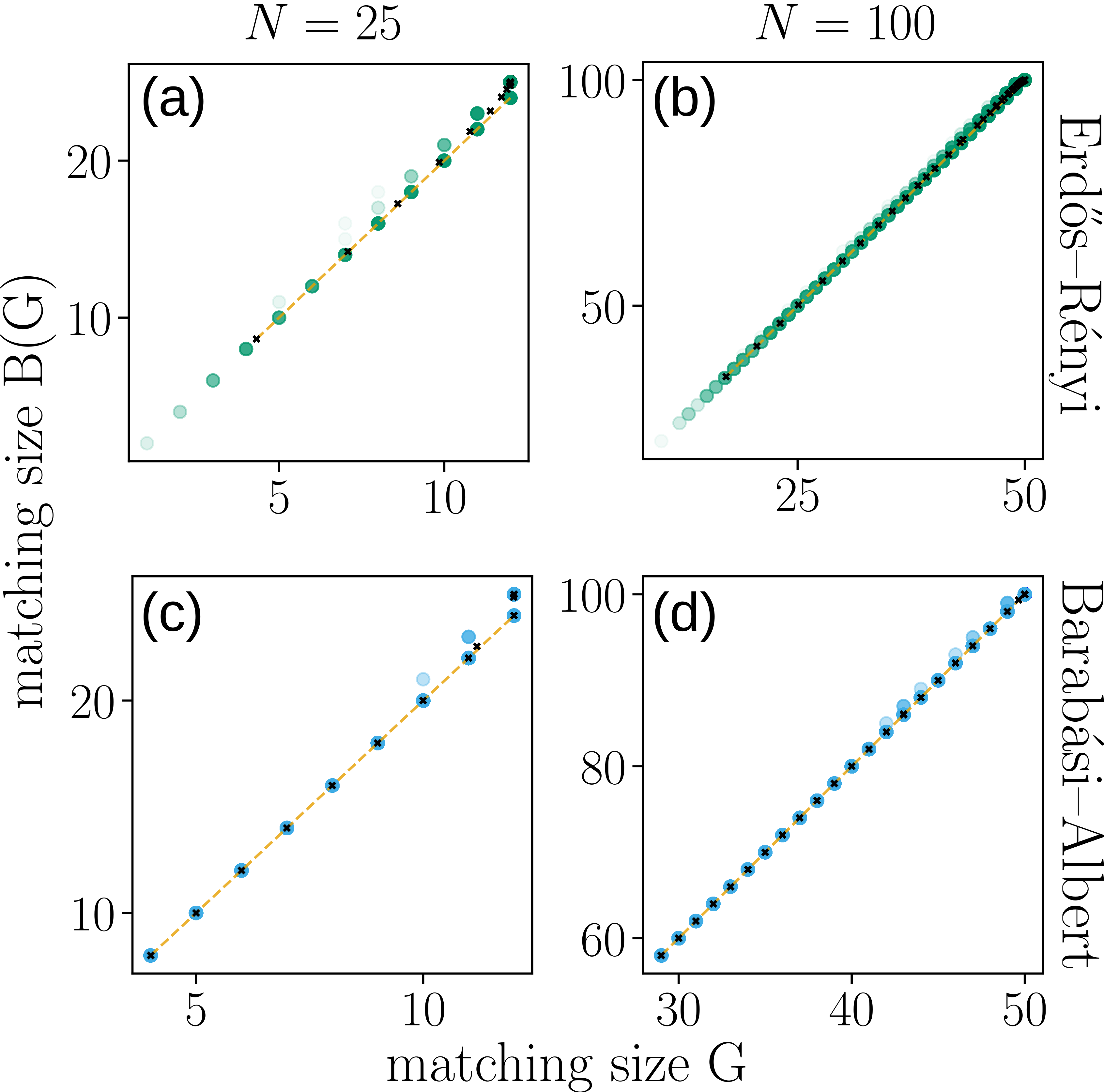}
    \caption{{\bf Maximum matchings in random graphs.}
    Panels (a) and (b) show the size of the maximum matching of $B(G)$ as a function of the size of the maximum matching of $G$ for ER graphs with $N=25$ and $N=100$ nodes respectively, for $20$ and $70$ different values of the edge creation probability parameter $p$ respectively.
    Each faint dot represents a single graph realization and the crosses depict the averages.
    Panels (c) and (d) show the same results for BA graphs with different attachment parameters $m$.
    The observed slope is approximately 2 in all cases, confirming that the bipartite double cover effectively doubles the matching size.}
    \label{fig:matching}
\end{figure}
In this section, we investigate the relationship between maximum matchings in a graph $G$ and its bipartite double cover $B(G)$.
We note in the main text that the lower bound for the maximum matching size of the bipartite double cover is $2\langle |M^*| \rangle$, where $M^*$ denotes a maximum matching of $G$.
This is immediate since every matching in $G$ canonically induces a matching of twice the size in $B(G)$.
However, there are cases in which the bipartite double cover admits an even larger matching. For example, consider the cyclic graph $G_3$ with three nodes.
While $G_3$ has a maximum matching of size $1$, the maximum matching of $B(G_3)$ has size $3$.
We therefore study the matching sizes of the bipartite double cover for the random graph models considered in the main text, and compare them to the maximum matching size of the underlying graph.
Figure~\ref{fig:matching} illustrates our findings for both ER and BA graphs with system sizes $N=25$ and $N=100$.
Each faint dot corresponds to a single graph realization, while the crosses denote the average over 200 realizations for each parameter value.
For ER graphs we vary the edge probability $p$, whereas for BA graphs we vary the attachment parameter $m$.
We confirm that twice the size of the maximum matching of the original graph is a lower bound for the maximum matching size of the bipartite double cover.
We also observe that this bound is typically, but not always, tight.
\end{document}